# Preregistration Does Not Improve the Transparent Evaluation of Severity in Popper's Philosophy of Science or When Deviations are Allowed

*Mark Rubin*
*Durham University, UK*

## Abstract

One justification for preregistering research hypotheses, methods, and analyses is that it improves the transparent evaluation of the severity of hypothesis tests. In this article, I consider two cases in which preregistration does not improve this evaluation. First, I argue that, although preregistration may facilitate the transparent evaluation of severity in Mayo's error statistical philosophy of science, it does not facilitate this evaluation in Popper's theory-centric approach. To illustrate, I show that associated concerns about Type I error rate inflation are only relevant in the error statistical approach and not in a theory-centric approach. Second, I argue that a test procedure that is preregistered but that also allows deviations in its implementation (i.e., "a plan, not a prison") does not provide a more transparent evaluation of Mayoian severity than a non-preregistered procedure. In particular, I argue that sample-based validity-enhancing deviations cause an unknown inflation of the test procedure's Type I error rate and, consequently, an unknown reduction in its capability to license inferences severely. I conclude that preregistration does not improve the transparent evaluation of severity (a) in Popper's philosophy of science or (b) in Mayo's approach when deviations are allowed.

Preregistration involves the time-stamped documentation of a study's planned hypotheses, methods, and analyses. The preregistered document is then made available with the final research report to allow others to identify any deviations from the planned approach (Nosek et al., 2019; Nosek et al., 2018).

Previous justifications for preregistration have argued that its primary role is to distinguish between "confirmatory" and "exploratory" analyses (e.g., Nosek & Lakens, 2014). However, there are unresolved questions about the meaningfulness of this distinction. For example, the distinction does not have an agreed formal definition in either statistical theory or the philosophy of science. In addition, critics have questioned related concerns about the "double use" of data and "circular reasoning" (Devezer et al., 2021; Rubin, 2020a, 2022; Rubin & Donkin, 2024; Szollosi & Donkin, 2021; see also Mayo, 1996, pp. 137, 271-275; Mayo, 2018, p. 319).

More recently, Lakens and colleagues have provided an alternative justification for preregistration based on Mayo's (1996, 2018) error statistical philosophy of science (Lakens, 2019, 2024; Lakens et al., 2024; see also Vize et al., 2024). In particular, Lakens (2019) argues that "preregistration has the goal to allow others to transparently evaluate the capacity of a test to falsify a prediction, or the severity of a test" (p. 221).

In this article, I consider two cases in which preregistration does not improve the transparent evaluation of severity. First, I highlight some differences between Mayo's (1996, 2018) error statistical account of severity and Popper's (1962, 1983) theory-centric account, and I argue that although preregistration may improve the transparent evaluation of Mayoian severity, it does not improve the transparent evaluation of Popperian severity. To illustrate my argument, I show that associated concerns about Type I error rate inflation are only relevant in an error statistical approach and not in a theory-centric approach.

Second, I argue that a preregistered test procedure that allows deviations (i.e., "a plan, not a prison"; Nosek et al., 2019, p. 817) does not provide a more transparent evaluation of Mayoian severity than a non-preregistered procedure. In particular, I argue that a test procedure that permits sample-based validity-enhancing deviations from its preregistered plan will include an unknown number of deviations during a hypothetical long run of its repetitions. These deviations cause an unknown inflation of the procedure's Type I error rate and, consequently, an unknown reduction of its capability to license inferences with Mayoian severity. I conclude that preregistration does not improve the transparent evaluation of severity (a) in Popper's philosophy of science or (b) in Mayo's approach when deviations are allowed.

## Preregistration Does Not Improve the Transparent Evaluation of Popperian Severity

In Mayo's (1996, 2018) error statistical approach, a hypothesis passes a severe test when there is a high probability that it would not have passed, or passed so well, if it was false (Mayo, 1996, p. 180; Mayo, 2018, p. 92; Mayo & Spanos, 2006, pp. 328, 350; Mayo & Spanos, 2010, p. 21; Mayo & Spanos, 2011, pp. 162, 164). Hence, severity is a characteristic of "the test *T*, a specific test result $x_0$, and a specific inference *H* (not necessarily predesignated)" (Mayo & Spanos, 2011, p. 164, italics occur in the original text for all quotes).

A test procedure's pre-data error probabilities play an important role in evaluating severity (Mayo & Spanos, 2006, p. 330).[1] In particular, "pre-data, the choices for the type I and II errors reflect the goal of ensuring the test is capable of licensing given inferences severely" (Mayo & Spanos, 2006, p. 350; see also Mayo & Spanos, 2011, p. 167). For example, a test procedure with a nominal pre-data Type I error rate of $\alpha = 0.05$ is capable of licensing specific inferences with a minimum "worst case" severity of 0.95 (i.e., 1 – α; Mayo, 1996, p. 399). Importantly, low error probabilities are necessary but not sufficient to license inferences severely (Mayo, 2018, pp. 13-14, 236, 396; Mayo & Spanos, 2011, p. 163). Error probabilities must also be relevant to the current inference (Mayo, 2018, pp. 194, 236, 429; Mayo & Spanos, 2006, p. 349), and statistical model assumptions must be approximately satisfied (Mayo, 2008, pp. 863-864; Mayo, 2018, p. 94; Mayo & Spanos, 2006, p. 349; Mayo & Spanos, 2011, pp. 189-190).

"Biasing selection effects" in the experimental testing context can lower the capability of a test procedure to license inferences severely by increasing the error probability with which the procedure passes hypotheses (Mayo, 2018, pp. 40, 196). For example, data dredging, fishing, cherry-picking, selective reporting, and *p*-hacking can represent biasing selection effects that increase a test procedure's error probability and, consequently, lower its capability for severe tests (e.g., Mayo, 1996, pp. 303-304; Mayo, 2008, pp. 874-875; Mayo, 2018, pp. 92, 274-275).

From an error statistical perspective, the goal of preregistration is to allow a more transparent evaluation of the capability of a test procedure to perform severe tests (e.g., Lakens, 2019; Lakens et al., 2024). In particular, preregistration reveals a researcher's *planned* hypotheses, methods, and analyses and enables a comparison with their *reported* hypotheses, methods, and analyses in order to identify any biasing selection effects in the experimental testing context that may increase the test procedure's error probabilities and lower its capability for severe tests.

Note that the more precisely specified a preregistered plan, the greater its potential to identify biasing selection effects. A vaguely specified preregistration that allows a lot of flexibility in the implementation of a planned test procedure has less potential to identify biasing selection effects and so will be less effective in allowing a transparent evaluation of the procedure's capability to perform severe tests (Lakens, 2019, pp. 226-227; Lakens et al., 2024, p. 16). Hence, it has been proposed that, ideally, preregistered research protocols should include machine-readable code that limits undisclosed analytical flexibility by automatically analyzing the data and evaluating the results (Lakens & DeBruine, 2021, pp. 10-11; Lakens et al., 2024, p. 16).

Importantly, Mayo's (1996, 2018) error statistical approach provides only *one* of *several* different conceptualizations of severity. Other conceptualizations have been proposed by Bandyopadhyay and colleagues (Bandyopadhyay & Brittan 2006; Bandyopadhyay et al., 2016), Hellman (1997, p. 198), Hitchcock and Sober (2004, pp. 23-25), Horwich (1982, p. 105), Lakatos (1968, p. 382), Laudan (1997, p. 314), Popper (1962, 1983), and van Dongen et al. (2023). Furthermore, preregistration may not improve the transparent evaluation of these other types of severity. In the present article, I illustrate this point by showing that preregistration may only improve the transparent evaluation of Mayoian severity, not Popperian severity.

I focus on Popperian severity because Popper's philosophy of science is relatively popular, and it has been used to support the case for preregistration. For example, Lakens et al. (2024) claimed that "preregistration is a methodological procedure that, given a specific philosophy of science (i.e., Popper's methodological falsificationism), improves one part of the research process – the evaluation of the test severity of hypothesis tests" (p. 5; see also Lakens, 2019, p. 227; Lakens, 2024, p. 1). Contrary to this view, I argue that Mayo's error statistical conceptualization of severity does not fit Popper's theory-centric philosophy of science (see also Mayo, 1996, pp. 240, 412; Mayo, 2018, p. 83), and that preregistration does not improve the evaluation of severity in Popper's approach. I begin by explaining Popperian severity and considering its differences with Mayoian severity.

## Popperian Severity

Popperian severity is measured as the conditional probability of a statement of supporting evidence (*e*) given the conjunction of a hypothesis (*h*) and background knowledge (*b*) divided by the conditional probability of *e* given *b* alone. In other words, severity (*e, h*, *b*) = *p*(*e*, *hb*)/*p*(*e, b*) (Popper, 1962, p. 391; Popper, 1966b, p. 288; see also Popper, 1983, pp. 238-239). Hence, the more probable is *e* given *hb* relative to *e* given *b* alone, the more severe the test of *h*. Note that "background knowledge" (*b*) refers to the initial conditions of a particular test together with relevant auxiliary hypotheses and theories that have been tentatively and temporarily accepted as being unproblematic during the test (Popper, 1962, pp. 238, 390; Popper, 1966b, p. 287).

As explained previously, a test procedure's capability to license inferences with Mayoian severity can be reduced by biasing selection effects in the experimental testing context. In particular, unplanned changes to the way in which a researcher constructs and selects hypotheses and evidence during the implementation of a test procedure may increase its error probability and decrease its capability for severe tests. Hence, the evaluation of Mayoian severity requires a consideration of the impact of any biasing selection effects (e.g., Mayo, 1996, pp. 303-304; Mayo, 2008, pp. 874-875; Mayo, 2018, pp. 274-275; Mayo & Cox, 2010, pp. 267-270). In contrast, the measurement of Popperian severity does not take account of the potentially biased way in which a researcher constructs or selects *e*, *h*, and/or *b* in the experimental testing context. As Mayo (1996) explained, in the case of Popperian severity, "there is no demand that a specific testing context be delineated, there are just…requirements in terms of the logical relationships between statements of evidence and hypotheses" (p. 209; see also Lakatos, 1978, p. 114; Mayo, 1996, pp. 206-207, 255, 330; Popper, 2002, pp. 64, 68, 267).

Popper would likely agree with Mayo's assessment. He argued that a researcher's private psychologically biased reasons for constructing and selecting particular *e, h*, and *b* for inclusion in a test (e.g., "because I want to get a significant result") belong to the subjective "World 2" context of discovery and, consequently, they do not enter into a deductive method of testing. In contrast, a researcher's public formal specification of a hypothesis test belongs to the autonomous objective "World 3" context of justification, which is open to logical and critical scrutiny (Popper, 1962, p. 140; Popper, 1974a, pp. 74, 118; Popper, 1983, p. 36; Popper, 1994, pp. 149-150; Popper, 2002, p. 7; see also Reichenbach, 1938, pp. 6-7).[2] Of course, a researcher's public World 3 approach may also be biased. However, as Popper (1994, p. 7) pointed out, "it need not create a great obstacle to science if the individual scientist is biased in favour of a pet theory," because science proceeds on the basis of the *collective* criticism of World 3's arguments and theories by other scientists (e.g., peer review, further tests, etc.; see also Dang & Bright, 2021). Hence, "if you are biased in favour of your pet theory, some of your friends and colleagues (or failing these, some workers of the next generation) will be eager to criticize your work – that is to say, to refute your pet theories if they can" (Popper, 1994, p. 93; see also Popper, 1966a, pp. 218-219). This collective "critical rationalist" approach is an essential part of Popper's philosophy of science (Popper, 1966a, pp. 229-231; Popper, 1994, p. 159), and it may be described as "theory-centric" because it occurs between and within relevant theories in World 3 (Popper, 1962, p. 26; Popper, 1974a, pp. 15, 82; Popper, 1983, pp. 28-30, 32; see also Musgrave, 2010).[3]

Although Popper was not concerned about *psychological* bias in researchers' construction and selection of hypotheses and evidence, he was concerned about an *epistemological* bias during testing. In particular, he argued that evidence *e* could only corroborate hypothesis *h* "if *e* is the result of genuine or sincere attempts to refute *h*" (Popper, 1983, p. 235; see also Popper, 2002, pp. 437-438). Note that *sincerity* "is not meant in a psychologistic sense" (Popper, 1974b, p. 1080).

Hence, it is not intended to address psychologically motivated biasing selection effects (Popper, 1974b, p. 1080). Instead, Popper’s “requirement of sincerity” (Popper, 2002, p. 437) represents a “methodological rule” (Popper, 1974b, p. 1080) that is intended to support his falsificationist epistemology over that of verificationism (Popper, 1983, p. 235). He proposed two ways of implementing this rule.

First, “we can partly formalize…[the requirement of sincerity] by demanding that our empirical test statements should be unexpected or improbable in the light of our background knowledge; that is to say, their probability, given the background knowledge, should be (considerably) less than ½” (Popper, 1983, p. 253). In other words, “$p(e, b) \ll \frac{1}{2}$” (Popper, 1983, p. 238). The more improbable is $e$ given $b$ alone, the more severe and sincere the test.

Second, we should design “crucial” tests in “which the theory to be tested predicts results which differ from results predicted by other significant theories, especially by those theories that have been so far accepted” (Popper, 1983, p. 235). In particular, we must pit our primary hypothesis $h$ against a significant, accepted, rival hypothesis $h'$ that predicts conflicting results given the same background knowledge (Popper, 1962, pp. 112, 197, 246; Popper, 1974a, pp. 13-15, 354, Footnote 7; Popper, 1974b, p. 995; Popper, 1983, pp. 188, 233-236; Popper, 1994, p. 7; Popper, 2002, p. 277; see also Bandyopadhyay & Brittan, 2006, p. 276; Bandyopadhyay et al., 2016, pp. 127-128; Lakatos, 1968, p. 380; Lakatos, 1978, p. 24, Footnote 1). It is only the refutation of $h'$ that allows a sincere corroboration of $h$ (Popper, 1974a, pp. 14-15; Popper, 1974b, pp. 995, 1009; Popper, 2002, pp. 66-67, 82).

In summary, Popper’s “requirement of sincerity” represents a methodological rule that we should “try to construct *severe* tests, and *crucial* test situations” (Popper, 1974a, p. 14). The more severe and crucial the test, the more it is “sincere,” and the less biased it is towards a “cheap” (verificationist) corroboration (Popper, 1983, pp. 130, 163, 257).

Popper argued that “the severity of our tests can be objectively compared; and if we like, we can define a measure of their severity” (Popper, 1962, p. 388; Popper, 1966b, p. 287; see also Popper, 1962, pp. 220, 390-391; Popper, 1983, pp. 238-239). However, he conceded that “the requirement of sincerity cannot be formalized” (Popper, 2002, p. 437), because “sincerity is not the kind of thing that lends itself to logical analysis” (Popper, 1983, p. 236; see also Popper, 1962, p. 288; Popper, 1983, pp. 244, 254; Popper, 2002, p. 419). Nonetheless, it remains possible to undertake an *informal* assessment of sincerity given the current state of World 3 knowledge (Popper, 1974b, p. 1080). In particular, collective critical rationalism may be used to evaluate the sincerity of a test by assessing the extent to which (a) $p(e, b) \ll \frac{1}{2}$ and (b) $h'$ represents a significant accepted theory that predicts contradictory results to $h$ (e.g., Lakatos, 1968, 1978; Laudan, 1997, pp. 314-315; see also Bandyopadhyay & Brittan, 2006, p. 264; van Dongen et al., 2023, p. 521). Hence, in Popper’s approach, an informal critical rational evaluation of sincerity can be used to contextualize and interpret a more formal measure of severity.[4] Importantly, and in contrast to Mayo’s error statistical approach, neither assessment requires a consideration of researchers’ private World 2 construction or selection of $e$, $b$, $h$, or $h'$ in the experimental testing context.

## Mayoian Severity

Mayo was not satisfied with Popper’s measure of severity (e.g., Mayo, 1996, p. 207; Mayo, 2006, p. 11), and she felt that his “theory-dominated” critical rational assessment of sincerity was inadequate (Mayo, 1997, p. 331; see also Mayo, 1996, pp. 59, 264; Mayo, 2006, p. 11; Mayo, 2018, pp. 40-41; Mayo & Spanos, 2006, p. 328). In her view, “it is impossible to assess reliability or severity with just statements of data and hypotheses divorced from the experimental context in which they were generated, modeled, and selected for testing" (Mayo, 2006, p. 36).

In response to these perceived deficiencies, Mayo (1996) developed an account of severity that refers to a test procedure's experimental testing context and its frequentist error probabilities across a hypothetical series of its repetitions (Mayo & Spanos, 2010, p. 21; see also Mayo, 1996, p. 72; Mayo, 2018, pp. 72-73; Mayo & Spanos, 2006, p. 328).[5] As she explained:

> "We must look at the particular experimental context in which the evidence was garnered and argue that its fitting a hypothesis is very improbable, if that hypothesis is false. This relativity to an experimental testing model and the focus on (frequentist) probabilities of test procedures distinguish my account, particularly from others that likewise appeal to probabilities to articulate the criterion for a good or severe test – even from accounts that at first blush look similar, most notably Popper's" (Mayo, 1996, pp. 206-207).

Importantly, Mayo's (1996) concept of an "experimental context" includes the potentially unreported and psychologically biased process by which researchers might construct and select hypotheses and evidence during the implementation of a test procedure. To be clear, like Popper, Mayo (1996) accepts that it does not matter that a researcher's psychologically biased intentions may influence the *specification* of a test procedure (Mayo, 1996, p. 409; see also Mayo, 1996, pp. 148, 263; Mayo, 2018, pp. 9-10). However, unlike Popper, she argues that it *does* matter that the researcher's psychological intentions may cause biasing selection effects during the *implementation* of that procedure. It matters, she argues, because biasing selection effects may increase the procedure's frequentist error probability and lower its capability to license inferences severely (e.g., Mayo, 1996, p. 349). Consequently, we must check or "audit" (Mayo, 2018, Tour III) the entire experimental testing context, including its unreported parts, in order to pick up on any biasing selection effects during the implementation of the test procedure and meet a minimal requirement for severity (Mayo, 1996, p. 298; Mayo, 2018, pp. 5, 9, 49, 92). In contrast, Popper (1967, pp. 34, 39) argued that hypothetical probability distributions are defined relative to a researcher's formal public World 3 specifications of an experiment and, consequently, they are not affected by unspecified (unreported and unknown) issues that may occur during the implementation of that experiment.

At a more general level, Mayo rejects Popper's "theory-dominated" approach and develops a philosophy of science in which "what we rely on…are not so much scientific theories but *methods* for producing experimental effects" (Mayo, 1996, p. 15; see also Mayo, 1996, pp. 11-12, 17 Footnote 3, 59; Mayo, 1997, p. 331). These experimental effects warrant only low-level local claims that are limited to specific experimental contexts (Mayo, 2006, pp. 37-38). The error statistical approach does not allow an individual test to logically refute universal, generalizable, theories or hypotheses (e.g., "all swans are white"; Chalmers, 2010, pp. 60-63; Musgrave, 2010, p. 6). Hence, the error statistical approach does not entail Popperian theory testing (see also Bandyopadhyay et al., 2016, p. 87), and it "does not find its home in a Popperian framework" (Mayo, 1996, p. 412). Instead, Mayo's approach is more consistent with the New Experimentalist view that "experiments, as Ian Hacking taught us, live lives of their own, apart from high level theorizing" (Mayo, 1996, pp. xiii, 12, 17, 190, 213; Musgrave, 2010, pp. 108-109). Again, this view conflicts with Popper's theory-centric approach in which "theory dominates the experimental work" (Popper, 2002, p. 90; cf. Mayo, 1996, pp. 59, 264; Mayo, 1997, p. 331; Mayo, 2006, p. 11;

Mayo, 2018, pp. 40-41; see also Popper's, 1983, pp. 47-48, 50, contrast between "inductivist" and theory-centric styles of reporting research).

Nonetheless, Mayo argues that the piecemeal testing of local, experiment-bound claims may inform decisions about higher-level theories (Mayo, 1996, p. 191; Mayo, 2010, pp. 35-36; Mayo & Spanos, 2010, p. 83). As she explained, "when enough is learned from piecemeal studies, severe tests of higher-level theories are possible" (Mayo, 1996, p. 191). Following a Bayesian conceptualization of severity, Bandyopadhyay et al. (2016, pp. 80-82) criticise Mayo's piecemeal approach, arguing that it falls foul of a "probability conjunction" error: If a higher-level theory is taken to pass a severe test only when all of its constituent local claims pass severe tests, then, following the conjunction rule in probability theory, the probability of observing all of the test results that support the theory will always be lower than the probability of observing a result that supports any single claim, with this discrepancy increasing as more claims are added. Hence, a piecemeal approach leads to the counterintuitive conclusion that, "the more local hypotheses that pass severe tests, the less probable on the data would be the global theory that comprises them" (Bandyopadhyay et al., 2016, p. 82). For a further discussion of this issue, please see Bandyopadhyay (2019).

Bandyopadhyay et al.'s (2016) criticism assumes that severity is based on likelihoods (Bandyopadhyay et al., 2016, p. 77). However, if severity is based on error probabilities, and all local claims must be passed in order to pass a higher-level theory, then, as the number of local claims increases, the probability that the theory would pass *if it were false* decreases, and so Mayoian severity increases. Mayo (2018) describes this approach as "an argument from coincidence to the absence of an error" (p. 16). It represents an informal, qualitative assessment of severity (Mayo, 1996, pp. 409-410; Mayo, 1997, p. 257; Mayo, 2006, p. 33; Mayo & Spanos, 2006, p. 348), and it is intended to solve the problem of induction (Mayo, 2018, p. 16). This issue illustrates how different conceptualisations of severity (i.e., based on likelihoods vs. error probabilities) can lead to different implications.*

## Illustrating the Differences Between Mayoian and Popperian Severity

To provide a more concrete illustration of the differences between Mayoian and Popperian severity, consider a researcher who conducts multiple uncorrected and unreported tests in order to find and report a single significant result (i.e., *p*-hacking). They then secretly hypothesize after their significant result is known in order to construct a hypothesis that they report as if it was generated before they conducted their analysis (i.e., HARKing). In this case, the researcher has been biased in the selection of their reported evidence (i.e., in favor of a test that yields a significant result) and the construction of their reported hypothesis (i.e., in favor of a hypothesis that is corroborated by that result). These biasing selection effects have made it easier for the researcher to pass false hypotheses during repetitions of their (partially unreported) test procedure. Consequently, their method has a relatively low capability to license inferences with Mayoian severity.

In contrast, these biasing selection effects do not affect Popperian severity or sincerity because they occur in the experimental testing context, and Popper's approach does not take

* This paragraph was not included in the published version of this article following an objection from an anonymous peer reviewer that Mayo (1996) also makes use of likelihoods. I note here that Mayo (1996, p. 394) was careful to distinguish between likelihoods and error probabilities in her conceptualization of severity (see also Mayo, 2018, pp. 30-32, 41). Hence, I think it is useful to consider her "argument from error" as a counterargument to Bandyopadhyay et al.'s (2016) criticism.

account of the experimental testing context (Mayo, 1996, p. 209). Popperian severity is measured as the conditional probability of a result (*e*) given the conjunction of a hypothesis (*h*) and background knowledge (*b*) relative to the probability of *e* given *b* per se. The researcher's privately biased selection of *e*, *h*, and/or *b* via *p*-hacking and HARKing does not enter into this measurement.

Popperian sincerity requires that (a) $p(e, b) \ll \frac{1}{2}$ and (b) $h'$ represents a significant accepted theory that predicts contradictory results to *h*. Certainly, this requirement may not be met, leading to a "cheap" (verificationist) corroboration (Popper, 1983, pp. 163, 257). However, the resulting bias is epistemological rather than psychological, and it can be evaluated via a critical rational discussion of publicly available information independent from any biasing selection effects that may be hidden in the experimental testing context (Popper, 1974b, p. 1080).

More generally, the researcher's private psychological reasons for conducting and reporting their specific test (i.e., because their previous tests did not yield a significant result) and for constructing their specific hypothesis (i.e., because it was passed by the current result) belong to Popper's World 2 of subjective intentions rather than his World 3 of objective specifications (Popper, 1974a, pp. 74, 108-109, 299). From a Popperian perspective, what is relevant is the researcher's formal, public, scientific specification of their hypothesis test, and they can rationally reconstruct this specification in World 3 in a way that is epistemically independent from their private psychological intentions in World 2 (Popper, 1974a, pp. 179, 242; see also Rubin, 2022, pp. 540-542; Rubin & Donkin, 2024, pp. 2023-2024).[6] Furthermore, the researcher's psychologically biased reasoning and behavior in World 2 does not imply an incorrect, invalid, or unsound hypothesis test in World 3. Consequently, in the Popperian approach, even unplanned and subjectively biased hypothesis tests may be objectively severe and sincere.

Mayo and Cox (2010, pp. 268, 271-272) provide a further example. A researcher intends to conduct a linear regression analysis of *y* on *x*. However, they make the post-data decision to conduct a regression of log *y* on log *x*. According to Mayo and Cox, if the researcher makes this decision because they conducted both tests and the second test provides "the more extreme statistical significance…, then adjustment for selection is required" (p. 271). On the other hand, if the researcher makes this decision because "inspection of the data suggests that it would be better to use the regression of log *y* on log *x*,…because the relation is more nearly linear or because secondary assumptions, such as constancy of error variance, are more nearly satisfied" (p. 268), then "no allowance for selection seems needed…[because] choosing the more empirically adequate specification gives reassurance that the calculated *p*-value is relevant for interpreting the evidence reliably" (p. 272). However, consider a situation in which the selected test provides *both* the more extreme statistical significance *and* the more empirically adequate specification. In this case, an error statistician should attempt to ascertain the researcher's subjective World 2 intentions because those intentions may indicate the operation of a biasing selection effect that alters the test procedure's capability to perform severe tests in its hypothetical repetitions (Mayo, 1996, pp. 348-349; Mayo, 2018, pp. 49, 286). Hence, critics have argued that error statisticians need to take account of private World 2 intentions "locked up in the scientist's head" (Mayo & Spanos, 2011, p. 186; see also Mayo, 1996, pp. 346-350; Mayo, 2008, pp. 860-861). In contrast, a theory-centrist would only refer to the researcher's public objective World 3 specifications because "science is part of world 3, and not of world 2" (Popper, 1974b, p. 1148). Consequently, following the principle of "*relativity to specification*" (Popper, 1967, p. 35), they would limit their inference to the formally reported test, including its "selected repeatable conditions" (Popper, 1983, p. 312) and its associated hypothetical sampling distribution (Fisher, 1955, p. 75; Fisher, 1956, pp. 29, 44, 77-78, 82; Rubin, 2024c).

Mayo (1996, 2018) would argue that Popper's approach is insufficient to address his requirement of sincerity in the above examples. As she explained:

> "If you engage in cherry picking, you are not 'sincerely trying,' as Popper puts it, to find flaws with claims, but instead you are finding evidence in favor of a well-fitting hypothesis that you deliberately construct – barred only if your intuitions say it's unbelievable. The job that was supposed to be accomplished by an account of statistics now has to be performed by *you*. Yet you are the one most likely to follow your preconceived opinions, biases, and pet theories" (Mayo, 2018, pp. 40-41).

However, from a Popperian perspective, even a "cherry-picked" hypothesis that is deliberately chosen because it is corroborated can be said to have undergone a sincere test as long as other researchers agree that (a) $p(e, b) \ll ½$ and (b) $h'$ represents a significant accepted theory that predicts contradictory results to $h$. The "cherry-picking" that concerned Popper occurs in a public, transparent manner at the epistemological level in World 3, rather than in a private, unreported manner at the psychological level in World 2.

Finally, it is important to reiterate that the Popperian requirement of sincerity is established via a public, theory-centric, critical rationalist appraisal given the current state of World 3 knowledge rather than via "your intuitions" (Mayo, 2018, p. 40). Of course, Mayo (2018, p. 41) is correct that this appraisal may be affected by "your preconceived opinions, biases, and pet theories." However, it is not only "performed by *you*" (Mayo, 2018, p. 40). Again, critical rationalism is a *collective* process in which researchers' opinions and biases are pitted against those of other researchers in an ongoing critical discussion in the scientific community (Popper, 1966a, pp. 217-219; Popper, 1974b, p. 1080; Popper, 1994, pp. 7, 93; Schaller, 2016, p. 111). As Popper (1974b) explained:

> "I have always tried to show that *sincerity in the subjective sense* is not required, thanks to the social character of science which has (so far, perhaps no further) guaranteed its objectivity. I have in mind what I have often called 'the friendly-hostile cooperation of scientists'" (p. 1080).

This friendly-hostile cooperation "does not require that scientists be unbiased, only that different scientists have different biases" (Hull, 1988, p. 22; see also Popper, 1962, pp. 14-18; Popper, 1966a, p. 219; Popper, 1994, pp. 22, 93). Hence, for Popper, an objective evaluation of the requirement of sincerity is not "supposed to be accomplished by an account of statistics" (Mayo, 2018, p. 40). It is supposed to be accomplished through a well-conducted critical discussion among scientists (Popper, 1966a, pp. 217-218; Popper, 1974a, p. 22; Popper, 1983, p. 48). As Popper (1983) explained, "objectivity is not the result of disinterested and unprejudiced observation. Objectivity, and also unbiased observation, are the result of criticism" (p. 48; see also Lakatos, 1978, p. 15; Popper, 1994, p. 93).

## Using Preregistration to Transparently Evaluate Severity

The differences between Mayoian and Popperian severity are particularly relevant in the context of preregistration. A well-specified preregistration may improve the transparent evaluation of Mayoian severity by allowing others to check for any biasing selection effects that have occurred in the otherwise hidden experimental testing context during the implementation of a test procedure (e.g., Lakens, 2019, 2024; Lakens et al., 2024; see also Mayo, 1996, p. 296; Mayo, 2018, p. 319; Staley, 2002, p. 289). However, the same rationale does not apply in the case of Popperian severity.

A valid measurement of Popperian severity can be made using a potentially *p*-hacked result (*e*), a potentially HARKed hypothesis (*h*), and potentially biased background knowledge (*b*). In addition, the requirement of sincerity can be transparently evaluated via a public, collective, critical rational discussion of $p(e, b)$ and $h$ vs. $h'$ given the current state of World 3 knowledge. Preregistration does not facilitate transparency in either case because neither evaluation requires knowledge of the researcher's planned hypothesis test or unreported biasing selection effects. Indeed, Popper's "theory-dominated" approach can be applied retrospectively via a rational reconstruction of an unplanned test based on the *e*, *b, h*, and $h'$ that we have to hand and in the context of the current state of scientific knowledge (Mayo, 1996, pp. 67-68; see also Lakatos, 1978, p. 114; Popper, 1967, pp. 35-36; Reichenbach, 1938, p. 5; cf. Mayo, 1996, p. 17).[7]

## Type I Error Rate Inflation

The differences between the error statistical and theory-centric approaches can also be illustrated in relation to Type I error rate inflation. The error statistical approach distinguishes between two types of Type I error rate. The "computed" error rate is based on the number of formally reported tests, whereas the "actual" error rate is based on the number of reported tests *and* unreported tests in the experimental testing context. If some tests are unreported, then the "actual" error rate will be higher than the "computed" error rate. Hence, uncorrected multiple testing and selectively reported significant results (i.e., *p*-hacking) may cause an inflation of the "actual" Type I (familywise) error rate above the "computed" error rate (Mayo, 1996, pp. 303-304; Mayo, 2008, pp. 874-875; Mayo, 2018, pp. 274-275; Mayo & Cox, 2010, pp. 267-270).

From this error statistical perspective, a well-specified preregistered plan is helpful because it can be used to transparently verify the number of tests in the planned experimental testing context, which may include some tests that were conducted during the implementation of the test procedure but not reported (Nosek et al., 2018, p. 2601; Nosek et al., 2019, p. 816). The "actual" number of tests (*k*), can then be used to compute the planned test procedure's "actual" familywise error rate (i.e., $1 - [1 - \alpha]^k$) and determine whether it is higher than the "computed" error rate.

For example, imagine that a researcher preregisters three tests (i.e., $k = 3$), each with a nominal alpha level of 0.05, but then selectively reports only one of these tests because it was the only one to yield a significant result (i.e., a biasing selection effect). In this case, the "actual" familywise error rate would be 0.14 ($1 - [1 - 0.05]^3$) even if the "computed" error rate for the reported test was 0.05 ($1 - [1 - 0.05]^1$). Correspondingly, the "actual" minimum level of severity ($1 - 0.14 = 0.86$) would be lower than the "computed" minimum level ($1 - 0.05 = 0.95$; Mayo, 1996, p. 399). Preregistration is useful in this type of situation because it allows the identification of Type I error rate inflation and the associated reduction in the test procedure's capability to perform severe tests.

Importantly, however, this error statistical perspective does not apply in a "theory first" approach (Rubin, 2024a). In this case, the "actual" Type I error rate does not necessarily refer to the "experiment-wide significance level" of the "entire experimental testing context" and its associated "experimental distribution" (Mayo, 2018, p. 275; Mayo, 1996, pp. 143, 298), because not all of the tests in the experimental testing context may be logically related to a reported statistical inference (Rubin, 2021b, 2024c). Instead, the "actual" (relevant) Type I error rate is the familywise error rate of the tests that are formally used to make a statistical inference about a (potentially unplanned) hypothesis, and the number of these tests ($k$) can be logically deduced from the formally reported statistical inference.

For example, if a statistical inference is made about a single *individual* null hypothesis $H_{0,1}$ based on a single test of that hypothesis (i.e., $k = 1$) using an α of 0.05, then the "actual" Type I error rate for that inference will be the same as the "computed" (reported) nominal error rate (0.05 or $1 - [1 - 0.05]^1$), even if the researcher performed multiple other planned or unplanned tests, secretly or transparently, during their implementation of the experiment (Hitchcock & Sober, 2004, pp. 23-25; Rubin, 2017a, 2021b, 2024b, 2024c). In this case, it would be illogical to argue that these other tests (e.g., tests of $H_{0,2}$, $H_{0,3}$, $H_{0,4}$, etc.) contribute to the error rate for the statistical inference about $H_{0,1}$ because they are not logically related to this inference, which is about $H_{0,1}$ per se.

Similarly, if an inference is made about a *joint* intersection null hypothesis that is composed of three constituent null hypotheses $\{H_{0,1}$ & $H_{0,2}$ & $H_{0,3}\}$ (i.e., $k = 3$), then the "actual" familywise error rate can be logically deduced from the formally reported statistical inference as being 0.14 (i.e., $1 - [1 - 0.05]^3$).[8] In this case, it would be illogical to use a familywise error rate that included a test of $H_{0,4}$, even if $H_{0,4}$ was planned and/or conducted, because a familywise error rate that includes $H_{0,4}$ warrants a different statistical inference to the one that is reported (i.e., an inference about $\{H_{0,1}$ & $H_{0,2}$ & $H_{0,3}$ & $H_{0,4}\}$ rather than $\{H_{0,1}$ & $H_{0,2}$ & $H_{0,3}\}$; Rubin, 2024b, 2024c).

Error statisticians might argue that theory-centrists are "using the wrong sampling distribution" in these examples because it does not reflect the "actual" test procedure (Spanos & Mayo, 2015, p. 3546). However, this argument depends on how we define the "actual" test procedure (Mayo, 1996, p. 304). In the error statistical approach, the "actual" procedure includes a researcher's unreported tests in the experimental testing context. For example, if a researcher tests null hypotheses $H_{0,1}$ and $H_{0,2}$ and then makes an inference about $H_{0,1}$ (because $p < 0.05$) but does not report their test of $H_{0,2}$ (because $p > 0.05$), then their "actual" test procedure includes the unreported test of $H_{0,2}$, and the "right" sampling distribution is given under the joint intersection null hypothesis $\{H_{0,1}$ & $H_{0,2}\}$ (i.e., the "global" or "universal" null; Mayo, 2008, p. 875; Mayo & Cox, 2010, p. 269; Mayo, 2018, p. 276). In contrast, in a theory-centric approach, the "actual" test procedure only includes tests that are logically related to the formally reported statistical inference. If that inference is restricted to $H_{0,1}$, then the test of $H_{0,2}$ is not part of the "actual" test procedure, and the "right" (relevant) sampling distribution is given under $H_{0,1}$, not $\{H_{0,1}$ & $H_{0,2}\}$ (Rubin, 2024b, 2024c; see also Popper, 1967, p. 36). Figure 1 illustrates this point.

In summary, in the error statistical approach, the "actual" Type I error rate is based on the tests in the experimental testing context, and it is inflated above the "computed" error rate when the computed error rate does not refer to all of these tests. Consequently, preregistration can be a useful way of revealing the planned testing context to identify any biasing selection effects, Type I error rate inflation, and corresponding reduction in a test procedure's capability to perform severe tests. In contrast, in a theory-centric approach, the "actual" Type I error rate can be logically deduced from the formally reported statistical inference, even if that inference was unplanned and selectively reported. In this case, any Type I error rate "inflation" will be the result of a logical inconsistency between a formally reported statistical inference and a formally computed familywise error rate rather than the result of undisclosed tests in the experimental testing context. This inconsistency can be identified and rectified through a logical analysis of the relation between the reported inference and error rate without needing to consult a preregistered plan (for examples, see Rubin, 2024b, 2024c).

**Figure 1**

*Illustration of Error Statistical and Theory-Centric Approaches to Type I Error Rates*

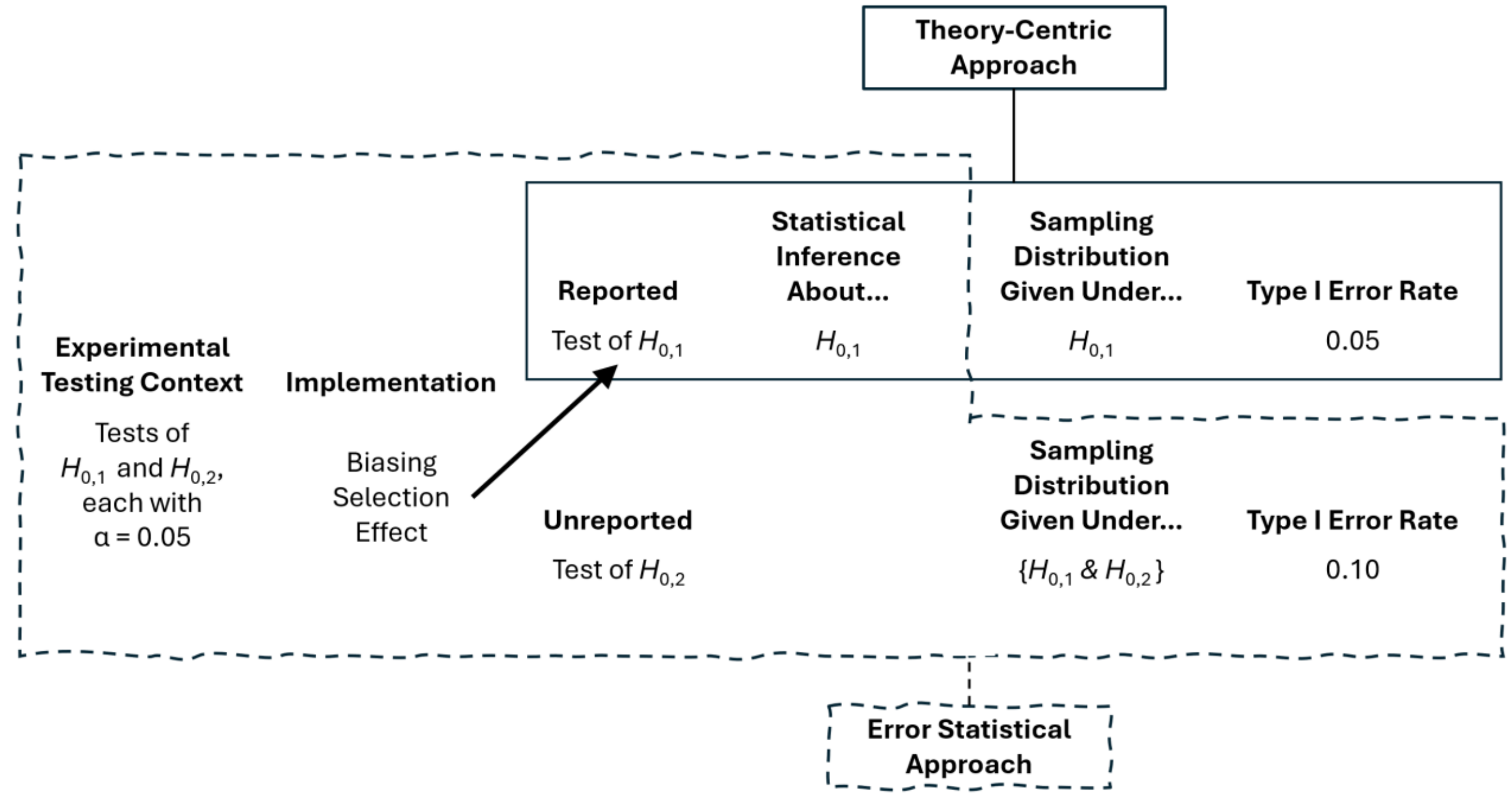

## Preregistration Does Not Improve the Transparent Evaluation of Mayoian Severity When Deviations are Allowed

So far, I have argued that preregistration does not improve the transparent evaluation of Popperian severity, but that a well-specified preregistration can improve the transparent evaluation of Mayoian severity. In this section, I add an important caveat to this argument: A well-specified preregistration can only improve the transparent evaluation of Mayoian severity *if the associated test procedure does not permit deviations in its implementation.* The deviations that I refer to here are those that are intended to maintain or increase the validity of a test procedure in light of unexpected issues that arise in particular samples of data. I argue that a test procedure that allows these sample-based validity-enhancing deviations during its implementation will suffer an unknown inflation of its Type I error rate and, consequently, an unknown reduction of its capability to license inferences with Mayoian severity.

### Sample-Based Validity-Enhancing Deviations Cause an Unknown Inflation of the Type I Error Rate

A researcher who preregisters their test procedure may encounter an unforeseen event or an unexpectedly violated assumption or falsified auxiliary hypothesis in their current sample. Assuming that they treat their preregistration as "a plan, not a prison" (Nosek et al., 2019, p. 817), they may then deviate from their preregistration in order to adapt their test procedure to maintain or increase its validity in light of this unanticipated issue (Nosek et al., 2018, p. 2602; Lakens, 2024, pp. 2, 7; Mayo & Cox, 2010, p. 268, Example 4; Rubin, 2017b, p. 326; Rubin & Donkin, 2024, p. 2030). For example, a researcher may adhere to a preregistered Student's *t*-test when the assumption of homogeneity is met in one sample. However, they may deviate from their plan and use Welch's $t$-test when this assumption is unexpectedly violated in another sample, because Welch's test provides a more valid approach in this situation (Lakens, 2024, p. 8). Consequently, in a hypothetical long run of random sampling, the researcher's test procedure would use *two* different tests to test the same joint intersection null hypothesis (i.e., Student's $t$-test & Welch's $t$-test). This "forking path" in the experimental testing context inflates the test procedure's "actual" Type I (familywise) error rate due to the multiple testing problem (Gelman & Loken, 2013, 2014; Rubin, 2017b; see also García-Pérez, 2012, pp. 4-5).

Note that sample-based validity-enhancing deviations do not represent Mayoian biasing selection effects because they are based on reasonable and/or conventional analytical principles with the aim of making a valid inference rather than a "desired inference" (Mayo, 2018, p. 105; see also Mayo & Cox, 2010, pp. 271-272). Nonetheless, during a hypothetical long run of repetitions of a test procedure that allows these deviations, the experimental testing context will include multiple tests that inflate the procedure's Type I (familywise) error rate. Hence, as Gelman and Loken (2013) explained, this multiplicity "can be a problem, even when there is no 'fishing expedition' or '*p*-hacking'" (p. 1; see also Rubin, 2017b, p. 324).

Also note that the forking paths issue is separate from the concern about the double use of data when checking test assumptions (e.g., when testing the assumption of homogeneity). Mayo is correct that, when researchers check test assumptions, the data can be considered to be "remodelled" to address a different question to the one addressed by the primary hypothesis (Mayo, 1996, pp. 137, 271-275; Mayo, 2018, p. 319). Hence, there is no circular reasoning in this case (see also Popper, 1962, p. 288; Popper, 1983, p. 133; Rubin & Donkin, 2024, p. 2023). Again, however, it remains the case that the introduction of a new test (e.g., Welch's $t$-test) based on sample-specific (data-dependent) information creates a forking path in the experimental testing context, and the uncorrected multiple testing in repetitions of the forked test procedure inflates its "actual" Type I (familywise) error rate.

A single sample-based validity-enhancing deviation opens up a single forking path in the experimental testing context. However, the error statistical approach operates on the basis of frequentist counterfactual reasoning that considers how a test procedure would perform given many different samples of data (Mayo, 2018, pp. 52-53). Consequently, during a hypothetical long run of random sampling, we must imagine that different samples may require different validity-enhancing deviations based on different unforeseen events, violated assumptions, and falsified auxiliary hypotheses. A test procedure that allows such deviations in its implementation will include an *unknown* number of deviation-based tests in its experimental testing context. Given that $k$ is unknown in this case, we cannot compute the test procedure's "actual" Type I (familywise) error rate using $1 - [1 - \alpha]^k$, and the error rate becomes "uncontrolled" (for related points, see Ditroilo et al., 2025, p. 1111; Nosek & Lakens, 2014, p. 138; Nosek et al., 2018, p. 2601; Nosek et al., 2019, p. 816; Rubin, 2017a, 2024c). Consequently, when a preregistered test procedure is treated as a "plan, not a prison," we cannot transparently evaluate the procedure's capability to license inferences with Mayoian severity (see also Mayo, 1996, pp. 313-314; Mayo, 2018, pp. 200-201; Staley, 2002, p. 289).

Contrary to this view, Lakens (2024) proposed that peers can evaluate whether sample-based validity-enhancing deviations increase or decrease Mayoian severity by considering (a) a researcher's flexibility with regards to other "plausible" and "defensible" analyses and (b) the results that follow from these alternative analyses (i.e., sensitivity analyses). Certainly, from a Popperian perspective, theories that allow a wider "range" ("Spielraum," "scope") of predictions in any given study will have lower "empirical content" and should therefore be downgraded a priori as being less "severely testable" (Popper, 2002, pp. 95, 108; see also Lakatos, 1968, pp. 375-376; Lakens, 2019, p. 224; Szollosi & Donkin, 2021, pp. 2-3; Rubin, 2017c, p. 316; Rubin, 2020a, p. 378; Rubin & Donkin, 2024, p. 2035). However, as discussed earlier, preregistration does not improve the evaluation of Popperian severity. Furthermore, it is important not to confuse Popperian and Mayoian severity here. From an error statistical perspective, if a test procedure allows sample-based validity-enhancing deviations in its implementation, then its capability for severe tests will be reduced by an unknown extent because each new sample may necessitate the addition of a new "plausible" and "defensible" test in the experimental testing context, leading to an unknown inflation of the procedure's Type I (familywise) error rate. Lakens (2024) does not consider this forking paths problem in his approach to sample-based deviations.

In summary, a test procedure that is preregistered but that also permits sample-based validity-enhancing deviations in its implementation will suffer an unknown inflation of its "actual" Type I (familywise) error rate and, consequently, an unknown reduction in its capacity to licence inferences with Mayoian severity. In this case, the only conclusion we can draw is that Mayoian severity is "low because we don't have a clue how to compute it!" (Mayo, 2018, p. 201; see also Mayo, 1996, p. 313; Mayo, 2018, p. 280), which is a conclusion that could also be reached in the absence of preregistration. Hence, if preregistration is treated as "a plan, not a prison," then it will not improve the transparent evaluation of Mayoian severity beyond that of a non-preregistered procedure (for related points, see Devezer et al., 2021, p. 17; Navarro, 2020, p. 8).

## Conditional Inference

One solution to the forking paths problem is to limit inferences to the single analytical path that has been followed in relation to the current sample (e.g., Welch's *t*-test) rather than the two paths that would be followed in a hypothetical long run of repeated random sampling (e.g., Student's *t*-test & Welch's *t*-test; Rubin, 2017b, p. 327; 2020a, 2024c). This *conditional inference* approach is consistent with Fisher's (1955, 1956) theory of significance testing, which refers to a hypothetical population that is assumed to represent the currently observed sample "in all relevant respects" (Fisher 1955, p. 72). To allow the opportunity for scientific progress and "learning by observational experience" (Fisher, 1956, pp. 99-100; see also Fisher, 1955, p. 73; Popper, 1983, pp. 40, 46), Fisher's approach also assumes that we do not yet fully understand all of the "relevant" aspects of the population, and that it contains undiscovered "relevant subsets" (subpopulations) that represent exceptions to the current conditional inference (Fisher, 1956, pp. 32–33, 55, 57, 80, 85-88; see also Popper, 1967, p. 39; Rubin, 2020b, 2021a). In this respect, Fisherian conditional inference is incompatible with the Neyman-Pearson theory of hypothesis testing, which assumes that Type I and II error rates apply *unconditionally* across a long run of random samples that are drawn from the same or equivalent fixed, fully-known, and well-specified population(s) (Fisher, 1955, p. 71; Rubin, 2021a, p. 5825). As Lehmann (1993) explained, this issue of conditional versus unconditional inference "seems to lie at the heart of the cases in which the two theories disagree on specific tests" (p. 1246).

In an attempt to bridge the gap between the Fisherian and Neyman-Pearson theories, Mayo (2014) proposed that we use a "weak conditionality principle" to condition Neyman-Pearson long-run error rates on "the experiment actually run" (p. 232; see also Mayo & Cox, 2010; Mayo, 2018, pp. 171-173). This weak conditionality principle is supposed to avoid the Type I error rate inflation caused by forking paths because it limits Neyman-Pearson error rates to tests that are "actually run" (e.g., Welch's *t*-test, not Student's *t*-test). However, unlike Fisher's single-sample version of conditionality, the weak conditionality principle continues to imply hypothetical repetitions of "the experiment actually run" and repeated random sampling from the associated population. Hence, consistent with the Neyman-Pearson approach, we must consider how "the experiment that *did* happen.…would behave in general, not just with these data, but with other possible data sets in the sample space" (Mayo, 2018, pp. 52-53). Furthermore, to be consistent with prior experience and a commitment to scientific progress, we must imagine that this experiment will sometimes encounter samples that necessitate deviations. Consequently, the weak conditionality principle does not prevent Neyman-Pearson error rates from becoming inflated to an unknown extent because "the experiment that *did* happen" is also prone to sample-based validity-enhancing deviations during a hypothetical long run of its repetitions.

To illustrate, imagine that a researcher deviates from a preregistered experiment $E_1$ in order to maintain or enhance the validity of their test following an unexpected event in the current sample. Based on the weak conditionality principle, they may exclude $E_1$ from their test procedure and condition their Neyman-Pearson long-run error rate on "the experiment actually run" (Mayo, 2014, p. 232), which we can denote as $E_2$ (e.g., Mayo & Cox, 2010, pp. 271-272, Example 4). Consistent with a theory-centric approach, this conditioning allows us to treat $E_2$ as an individual test rather than a union-intersection test of a "mixture experiment" (i.e., $\{E_1$ & $E_2\}$; Mayo, 2014, p. 228; Rubin, 2021b, 10973-10974). However, in this case, the researcher must also imagine that, as per their experience with $E_1$, a long run of repetitions of $E_2$ would sometimes encounter random samples that necessitate further validity-enhancing deviations. They may approach this fractal forking paths problem in one of two ways.

First, the researcher may prohibit any further sample-based validity-enhancing deviations during $E_2$'s implementation. In other words, they may permanently fix the current specification of $E_2$. This approach would preserve the applicability of $E_2$'s conditional long-run error rate. However, it would contradict the researcher's previous "plan, not a prison" rationale for deviating from their preregistration of $E_1$ to conduct $E_2$. Furthermore, permanently fixing $E_2$'s current specification implies that it will lack validity in relation to some of the samples that it encounters during its hypothetical repetitions (e.g., Lakens, 2024, p. 5; Rubin, 2017b, p. 326; Rubin & Donkin, 2024, p. 2030; see also García-Pérez, 2012, p. 5). Mayoian severity requires that a test procedure's statistical assumptions are approximately satisfied during testing because violated assumptions may inflate the procedure's "actual" error probability (Mayo, 2008, pp. 863-864; Mayo, 2018, p. 94; Mayo & Spanos, 2006, p. 349; Mayo & Spanos, 2011, pp. 189-190). $E_2$'s fixed test procedure is designed to fail this criterion because it cannot be modified when statistical assumptions are violated in particular samples. Consequently, consistent with the error statistical approach's frequentist counterfactual reasoning (Mayo, 2018, pp. 52-53), $E_2$ will have a low capability to license specific inferences severely in its hypothetical repetitions.

Second, the researcher may continue to allow sample-based validity-enhancing deviations of $E_2$'s test procedure. However, as discussed previously, this approach will open up an unknown number of forking paths in the experimental testing context, with each path specifying a new experimental test (i.e., $E_3, E_4, \ldots E_k$). $E_2$'s "computed" conditional Neyman-Pearson error rate will not apply across this "garden of forking paths" (Gelman & Loken, 2013; i.e., the mixture experiment $\{E_2$ & $E_3$ & $E_4$ & … $E_k\}$). Furthermore, it will not be possible to compute an "actual" unconditional Neyman-Pearson error rate for the test procedure because the number of tests in the experimental testing context (*k*) will be unknown (for related points, see Mayo, 1996, p. 317; Mayo, 2018, p. 39). Figure 2 illustrates this situation.

**Figure 2**

*Illustration of the Application of the Weak Conditionality Principle*

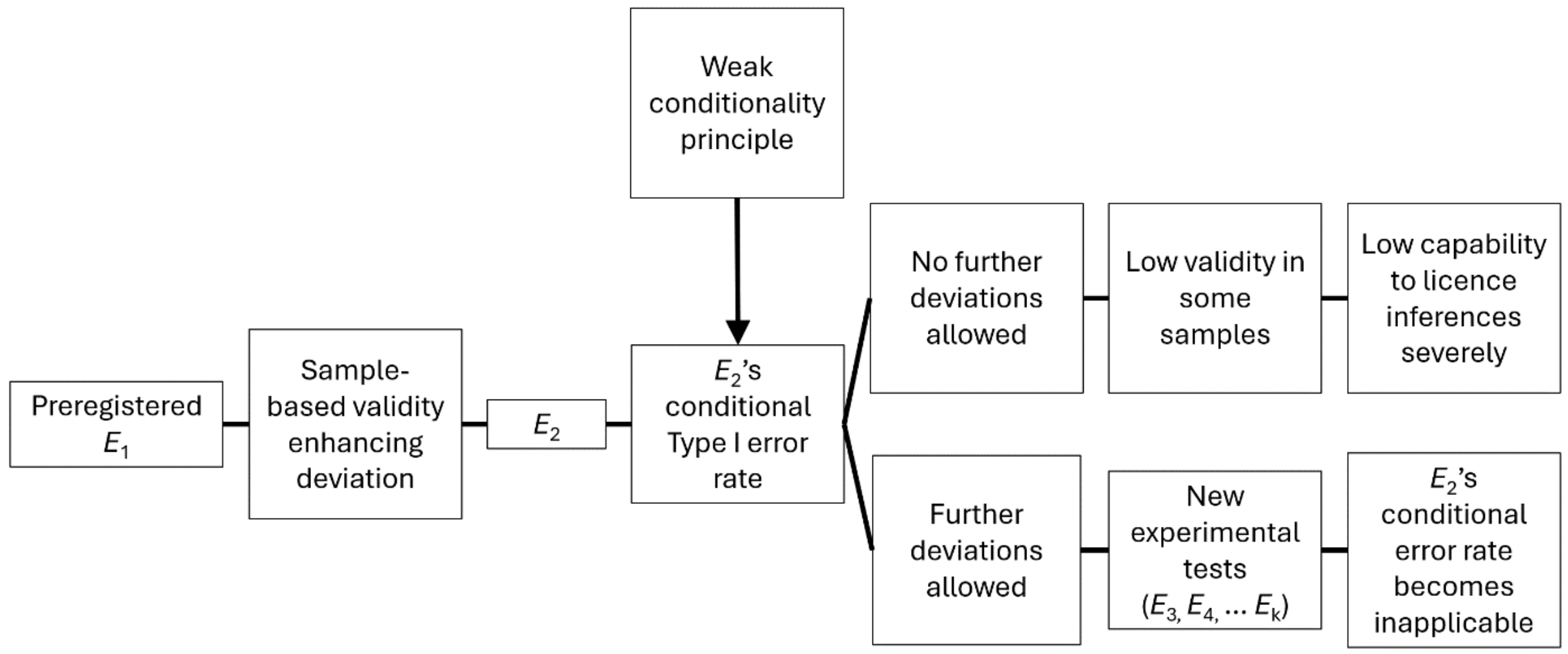

In summary, the decision to deviate from a preregistration often entails a trade-off between the goals of validity and error control (see also Lakens, 2024, p. 5). If a test procedure disallows sample-based validity-enhancing deviations in its implementation, then it will lack validity in some samples, and if it allows them, then its "computed" conditional error rate will become inapplicable and it will suffer an unknown inflation of its "actual" unconditional error rate. Mayoian severity will be compromised in both cases because it requires a method to be both valid and reliable (Mayo, 2008, pp. 863-864; Mayo & Spanos, 2006, pp. 349-350; Mayo & Spanos, 2011, p. 167).

## Do Researchers Treat Preregistration as a Plan, Not a Prison?

My argument implies that even a well-specified preregistration that results in no deviations when testing a particular sample will not improve the transparent evaluation of Mayoian severity if it is treated as an adjustable plan, rather than a fixed test procedure, because adjustable plans do not control the Type I error rate during a hypothetical long run of random sampling. However, the question remains as to how likely it is for researchers to treat their preregistrations as plans rather than prisons. This question can be addressed descriptively and normatively.

From a descriptive perspective, there is evidence that, in practice, researchers from several different fields tend to treat preregistrations as adjustable plans rather than fixed procedures because they deviate from those preregistrations. For example, in psychology, the percentage of studies that include at least one deviation (disclosed or undisclosed) ranges from 59% (Cheng, 2022; $N$ = 98 studies) to 93% (Claesen et al., 2021; $N$ = 27 studies). In neuroscience, the estimate is 84% (Clayson et al., 2025; $N$ = 92 articles); in exercise oncology trials, it is 87% (Singh et al., 2021; $N$ = 31); and in gambling studies it is 65% (Heirene et al., 2024; $N$ = 20, undisclosed deviations only). Finally, a meta-analysis of articles that looked at 1,113 (mainly) clinical trials found 41%-75% of studies with at least one outcome discrepancy (TARG Meta-Research Group and Collaborators, 2023). On average, across these five fields, around three-quarters of preregistered studies included at least one deviation, suggesting that, in practice, researchers regard preregistrations as adjustable plans, rather than fixed test procedures.

From a normative perspective, the above evidence is unsurprising given that leading proponents of preregistration encourage researchers to treat preregistration as "a plan, not a prison" (Chambers, 2019, pp. 188, 189; DeHaven, 2017; Nosek et al., 2019, p. 817). As discussed above, however, this approach makes preregistration ineffective for transparently evaluating Mayoian severity because, following the error statistical approach's frequentist counterfactual reasoning (Mayo, 2018, pp. 52-53), adjustable plans do not control the Type I error rate (see also Navarro, 2020, p. 8).

## When is Preregistration Useful and When is it Detrimental?

So, when is preregistration useful, and when is it detrimental? According to my argument, a well-specified preregistration will improve the transparent evaluation of Mayoian severity when it is treated as a fixed test procedure rather than an adjustable plan (see also Staley, 2002, p. 289). However, test users are unlikely to treat preregistrations in this way unless they are confident that they will not encounter any unforeseen events, assumption violations, or falsified auxiliary hypotheses in any of their samples. In other words, for preregistration to be useful for evaluating Mayoian severity, test users must have a high degree of confidence about the applicability of the background knowledge that underlies their test, perhaps as a result of extensive assumption testing. This level of confidence might occur in areas such as quality control testing during industrial production (Fisher 1955, pp. 69–70; Fisher 1956, pp. 99–100; Rubin, 2020b). Here, test users know the relevant and irrelevant aspects of their tests, the smallest effect size of interest, and the actual costs of Type I and II errors. However, as Popper (1962) explained, this high degree of confidence is inappropriate in scientific contexts, where researchers do "not *accept*…background knowledge; neither as established nor as fairly certain, nor yet as probable…[and they know] that even its tentative acceptance is risky, and…that every bit of it is open to criticism" (p. 238). Failure to fully embrace this critical attitude to background knowledge leads to *naïve methodological falsificationism* (Lakatos, 1978, p. 42), which Popper (1983, pp. xxii-xxiii, xxxv) rejected.

Preregistration may even be *detrimental* in scientific contexts. I consider two possibilities here. First, the preregistration of a test procedure may increase a researcher's commitment to that procedure relative to a situation in which it was not preregistered (Rubin & Donkin, 2024, p. 227). In turn, this *researcher commitment bias* may unconsciously deter the researcher from (a) considering high-quality criticisms of their preregistered approach, (b) deviating from their preregistration to adopt more valid tests, and (c) exploring their data to discover more informative results.[9]

Second, preregistration may add unwarranted epistemic value to a prediction that owes its success to a researcher's atheoretical guesswork rather than a theory's predictive power (Rubin & Donkin, 2024, p. 2028; for related work, see Grüning & Mata, 2024). Transparently hypothesising after the results are known (THARKing; Hollenbeck & Wright, 2017) serves to eliminate this *researcher prophecy bias* (Rubin & Donkin, 2024).

## Summary and Conclusion

Preregistration represents a somewhat contentious solution to an ill-defined problem. Previous justifications for this research practice have focused on the distinction between "exploratory" and "confirmatory" research. However, this distinction lacks an agreed statistical or philosophical basis (Devezer et al., 2021; Rubin, 2020a, 2022; Rubin & Donkin, 2024; Szollosi & Donkin, 2021).

Lakens and colleagues provide a more coherent justification based on Mayo's (1996, 2018) error statistical approach (Lakens, 2019, 2024; Lakens et al., 2024; see also Vize et al., 2024). From this perspective, the goal of preregistration is to allow others to transparently evaluate the capability of a test procedure to license inferences severely.

However, Mayo's (1996, 2018) error statistical conceptualization of severity is only one of several conceptualizations (Bandyopadhyay & Brittan, 2006; Bandyopadhyay et al., 2016; Hellman, 1997; Hitchcock & Sober, 2004; Horwich, 1982; Lakatos, 1968; Laudan, 1997; Popper, 1962, 1983; van Dongen et al., 2023). In the present article, I focused on Popper's conceptualization and showed that preregistration does not improve the transparent evaluation of either severity or sincerity in his theory-centric approach.

It is possible that a consideration of other approaches to severity may also reveal the redundancy of preregistration. For example, preregistration may not improve the transparent evaluation of Bayesian formulations of severity (e.g., Bandyopadhyay & Brittan, 2006; Bandyopadhyay et al., 2016; van Dongen et al., 2023) given that prior probability distributions are transparent (Rubin, 2022, pp. 540-542). Future work should consider this issue in greater depth.

I also showed that preregistration does not improve the transparent evaluation of Mayoian severity when deviations are allowed. In particular, I argued that a test procedure that is preregistered but that allows sample-based validity-enhancing deviations in its implementation will have an unknown inflation of its Type I (familywise) error rate and an unknown reduction of its capability to license inferences with Mayoian severity. Consequently, if preregistration is treated as only a plan, and not a prison, then it will not improve the transparent evaluation of Mayoian severity beyond that of non-preregistered research because potential deviations from that plan will inflate the procedure's "actual" Type I error rate by an unknown extent.

In conclusion, Mayo's (1996, 2018) error statistical approach justifies the use of a well-specified preregistered research plan to allow others to evaluate the capability of a fixed test procedure to license inferences severely. However, preregistration does not improve the transparent evaluation of severity in Popper's philosophy of science or in Mayo's approach when deviations are allowed.

# Endnotes

1. The error statistical approach also considers post-data error probabilities based on observed *p* values (Mayo, 2018, p. 440; Mayo & Spanos, 2006, pp. 333-334). Consistent with previous work in this area (Lakens, 2019; Lakens et al., 2024), I focus on pre-data error probabilities, and Type I error rates in particular, because they provide the clearest justification for the use of preregistration (see also Ditroilo et al., 2025, p. 1109).
2. Popper (1974a, pp. 74, 108-109) distinguished between three worlds. World 1 is the physical world. World 2 is the private psychological world of subjective beliefs, intentions, and experiences. Finally, World 3 is the public epistemological world of objective problems, theories, and reasons. Similar to Worlds 2 and 3, Reichenbach (1938, pp. 6-7) distinguished between (a) "processes of thinking in their actual occurrence" in the psychological "context of discovery" and (b) the "rational reconstruction" or "logical substitute" of these subjective thought processes in an epistemological "context of justification."
3. The term "theory-centric" is intended to denote a consideration of theories as the starting point for any scientific investigation. This consideration is not necessarily impartial (Popper, 1962, pp. 14-18; Popper, 1966a, p. 219; Popper, 1994, pp. 22, 93), and it does not necessarily refer to high-level theories such as the theory of relativity. It may also refer to simple theories such as "all swans are white" (e.g., Popper, 1983, p. xx) and background theories such as auxiliary theories about measurement (Popper, 1962, pp. 238, 390; Popper, 1966b, p. 287).

4. Popper (1962, pp. 247-248) argued that, for scientific knowledge to grow, theories must replace one another based on their successful predictions of new effects in crucial tests. However, determining which effects are "new" (i.e., $p[e, b] \ll ½$) and which tests are "crucial" ($h$ vs. $h'$) requires a conjectural reconstruction of problem-situations within their original historical contexts (see also Popper, 1974a, p. 176). There are two points to note here. First, different conjectural reconstructions that refer to different historical contexts will yield different conclusions (Lakatos, 1968, p. 387; Lakatos, 1978, pp. 79, 86). Hence, contrary to the error statistical approach, judgements of Popperian sincerity may vary depending on their historical context (Chalmers, 2010, p. 60). Second, conjectural reconstructions in World 3 do not require a consideration of a researcher's subjective experiences in World 2. Indeed, a researcher's conscious experience of a problem-situation may be quite different from a conjectural reconstruction of that situation (Popper, 1974a, pp. 179, 242).
5. Following the Neyman-Pearson approach to statistical hypothesis testing, Mayo (1996) conceptualizes error probabilities as occurring "in a long series of trials of this experiment" (p. 181; see also Mayo & Spanos, 2011, p. 162). The problem with this approach is that it assumes that we know the theoretically relevant and irrelevant aspects of "this experiment" (Staley, 2002, pp. 288-289). Mayo (1996, pp. 172-173, 298-299) did not consider this *reference class problem* to be a difficulty for her approach. However, Fisher (1955, p. 71; 1956, pp. 77-78, 82, 91) saw it as a fatal flaw in the scientific application of the Neyman-Pearson approach because, as scientists, we must concede that we do not fully understand the relevant and irrelevant aspects of our experimental procedures (Rubin, 2020b; Schaller, 2016, p. 108). Popper would agree that our experimental procedures are "impregnated" with fallible theories (Popper, 1974a, p. 145; Popper, 1983, p. 312; Popper, 2002, p. 94; see also Lakatos, 1978, p. 54; Popper, 1962, pp. 230, 238). As he noted, "any single case has so many properties that we cannot say, just by inspection, which of them are to be included among the specifications defining what should be taken as 'our' experiment, and as 'its' repetition" (Popper, 1967, pp. 38-39; see also Popper, 1983, p. 308). We may unintentionally exclude theoretically important specifications (Popper, 1962, p. 230). Hence, we must imagine that "a long series of trials of this experiment" will contain currently unknown, theoretically relevant variations of "this experiment" (i.e., relevant subsets differentiated by hidden moderators) that imply different error probabilities (Fisher, 1955, p. 71; Fisher, 1956, p. 33).
6. Popper (1974b) also argued that researcher deception does not pose a fundamental difficulty for science. He noted that even researchers who believe they are reporting the truth may, in fact, be making objectively false statements (see also Dang & Bright, 2021, p. 8196). Consequently, science must always proceed via questioning, critical discussion, and decision making rather than by attempting to discern whether particular researchers are reporting what they believe to be the truth. As Popper (1974b) explained, "I assert that, however different, psychologically, lying may be from speaking what is subjectively felt to be the truth, it is not this psychological difference that science is ultimately based on, but the critical tests by which we try to discern the objective difference between truth and falsity" (p. 1113).
7. Popper argued that "a hypothesis can only be empirically *tested…after* it has been advanced" (Popper, 2002, p. 7). However, he made this point to distinguish between inductivism and deductivism rather than the pre- and post-designation of specific tests, and predictions (basic statements) can be deduced from hypotheses and background knowledge *after* viewing corroborating or refuting results (i.e., "retrodiction" & "explicanda"; Popper, 2002, p. 38, Footnote *2; see also Brush, 2015, p. 78; Lakatos, 1978, pp. 35, 72-73, 114, 185; Rubin & Donkin, 2024). Popper also argued that a hypothesis test requires agreement among investigators about what constitutes (a) unproblematic background knowledge (Popper, 1962, p. 237; Popper, 1983, p. 244), (b) an intersubjectively testable experiment (Popper, 1974b, p. 970; Popper, 2002, p. 63, 86), and (c) "*criteria of refutation*…laid down beforehand" (Popper, 1962, p. 38, Footnote 3). However, this agreement is temporary, tentative, and open to challenge and revision "at any time" (Popper, 1962, p. 238; Popper, 1974a, p. 34; Popper, 1994, p. 160; see also Lakatos, 1978, pp. 42-45; Popper, 2002, p. 63). Consequently, pre-data agreements can be criticized and superseded by post-data agreements that incorporate new information and better interpretations (Popper, 1983, pp. xxx, Footnote 10; 188-189; 189, Footnote 3; Popper, 2002, p. 63; see also Lakatos, 1978, p. 45), even if one or more parties to the agreement are guided by hidden (World 2) agendas. Only *naïve methodological falsificationism* prohibits this revisionism (Lakatos, 1978, p. 42), and Popper (1983, pp. xxii-xxiii, xxxv) rejected this naïve approach because it is divorced from collective critical rationalism.

8. Joint hypotheses are indicated within braces { } throughout. Also note that, like individual hypotheses, joint hypotheses are *statistical* hypotheses (study-specific predictions) rather than *substantive* hypotheses (universal statements). This point is important because Type I error rates only refer to inferences about statistical hypotheses, not substantive hypotheses. In particular, they do not cover the wide range of inferential errors that may occur over and above random sampling error (Rubin, 2024c, p. 49).
9. Lakens et al. (2024, p. 13) appear to characterize the researcher commitment bias as a *conscious* bias. However, the various cognitive biases that are thought to underlie the researcher commitment bias are not restricted to conscious reasoning (i.e., the automation bias, plan continuation bias, commitment bias, and first-is-best bias; Rubin & Donkin, 2024, p. 227). Hence, like many other researcher biases, the researcher commitment bias may operate at an unconscious level. Lakens et al. (2024) go on to argue that the best way to mitigate the researcher commitment bias is through education and reporting templates. However, it is unclear whether this approach is sufficient to address an unconscious bias that, like the confirmation and hindsight biases, may fall into a "bias blind spot" (Pronin & Hazel, 2023; see also Nosek et al., 2018, p. 2601).

---

*Acknowledgments*: I am grateful to Deborah Mayo for her comments on a previous version of this article. Please note that she does not agree with all of the points I have made.

*Funding:* I declare no funding sources.

*Conflict of interest:* I declare no conflict of interest.

*Biography*: I am a professor of psychology at Durham University, UK. For further information about my work in this area, please visit https://sites.google.com/site/markrubinsocialpsychresearch/replication-crisis

*Correspondence:* Correspondence should be addressed to Mark Rubin at the Department of Psychology, Durham University, South Road, Durham, DH1 3LE, UK. E-mail: Mark-Rubin@outlook.com

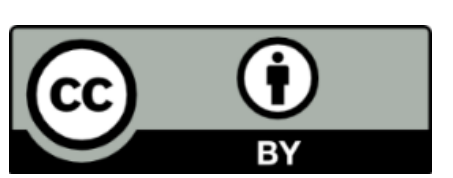

---